\documentclass[preprintnumbers,superscriptaddress,pra,showkeys]{revtex4}
\usepackage{amsfonts}
\usepackage{amssymb}
\usepackage{amsmath}
\usepackage{epsfig}
\usepackage{graphicx}
\usepackage{color}

\input{tcilatex}
\begin{document}

\title{Breakdown of equipartition of energy for vibrational heat capacity of
diatomic molecular gas due to nonvanishing bond length }
\author{X. Y. Wu}
\affiliation{School for Theoretical Physics, School of Physics and Electronics, Hunan
University, Changsha 410082, China}
\author{S. Y. Wang}
\affiliation{School for Theoretical Physics, School of Physics and Electronics, Hunan
University, Changsha 410082, China}
\author{J. R. Yang}
\affiliation{School for Theoretical Physics, School of Physics and Electronics, Hunan
University, Changsha 410082, China}
\author{X. Wang}
\affiliation{School for Theoretical Physics, School of Physics and Electronics, Hunan
University, Changsha 410082, China}
\author{Q. H. Liu}
\email{quanhuiliu@gmail.com}
\affiliation{School for Theoretical Physics, School of Physics and Electronics, Hunan
University, Changsha 410082, China}
\date{\today }

\begin{abstract}
When the theorem of equipartition of energy applies to the vibrational
degree of freedom within diatomic molecular gas, the bond length is usually
taken as zero so that the theorem is valid. Once the bond length is taken
into consideration, calculations show that the mean energy of the
vibrational heat capacity will significantly deviate from the standard value
near the high temperature which breaks the bond.
\end{abstract}

\maketitle

The theorem of equipartition of energy plays a crucial role in Boltzmann
statistical mechanics or classical statistical mechanics, which states that
molecules in thermal equilibrium have the same average energy associated
with each independent degree of freedom of their motion and that the energy
is $kT/2$, where $k$ is the Boltzmann constant and $T$ denotes the absolute
temperature. For an ideal monatomic gas, the mean energy per molecule is $%
3kT/2$ for there are only three translational degrees of freedom. For an
ideal diatomic molecular gas such $H_{2}$, $O_{2}$ and $CO$ etc., there is a
vibrational degree of freedom in which there is $kT/2$ coming from kinetic
energy and $kT/2$ from potential energy, so it is usually claimed that the
mean energy per molecule for each\ vibrational degree of freedom is $kT$ 
\cite{1,2,3,4,5}. For a diatomic molecule, we take the harmonic
approximation of the interaction between two atoms, and then we find that
relative to one atom, another atom oscillates with a frequency around an
equilibrium point, thus defining a bond length. For all diatomic molecules,
the bond length between two atoms are finite, for instance, for $H_{2}$, $%
O_{2}$ and $CO$ gas, it is $74$, $148$, $143$ picometers, respectively \cite%
{table}. To note that the so-called bond energy is characterized by $m\left(
\omega r_{0}\right) ^{2}/2$, where $m$ is the reduced mass, $\omega $ is the
vibrational frequency and $r_{0}$ is the bond length. An important situation
appears at dissociation temperature\ $T_{d}$ determined by $kT_{d}\approx
m\left( \omega r_{0}\right) ^{2}/2$; the bond breaks and the diatomic
molecule will be dissociated. Once the bond length is effectively taken as
zero, the theorem of equipartition of energy is valid.

The present study demonstrates how equipartition of energy for vibrational
potential energy with a nonvanishing bond length can be seriously deviated
from $kT/2$ at high temperature. For simplicity, we confine ourselves to an
ideal model of the diatomic molecule without considering the non-harmonic
oscillations, nor accounting for the non-separability between rotational and
vibrational degrees of freedom of the molecule near dissociation
temperature\ $T_{d}$.

Partition function for vibrational potential energy within a diatomic
molecule gas is usually given by,%
\begin{equation}
Z_{1}^{vp}=\int_{-\infty }^{\infty }\exp (-\frac{1}{2}\beta m\omega
^{2}r^{2})dr=\sqrt{\frac{\pi }{2\beta m\omega ^{2}}},  \label{1}
\end{equation}%
where $\beta =1/kT$. The \emph{mean vibrational potential energy} $u^{vp}$
per molecule is, 
\begin{equation}
u^{vp}=-\frac{\partial }{\partial \beta }\ln Z_{1}^{vp}=\frac{1}{2}kT.
\end{equation}%
It is right the theorem of equipartition of energy for the vibrational
potential energy.

Now, we take the finite bond length $r_{0}$ into consideration, we have the
partition function for the \emph{vibrational potential energy},%
\begin{equation}
Z_{1}^{vp}=\int_{0}^{\infty }\exp (-\frac{1}{2}\beta m\omega ^{2}\left(
r-r_{0}\right) ^{2})dr=\frac{\sqrt{\frac{\pi }{2}}\left( \func{erf}\left( 
\frac{r_{0}\sqrt{\beta m\omega ^{2}}}{\sqrt{2}}\right) +1\right) }{\sqrt{%
\beta m\omega ^{2}}},  \label{3}
\end{equation}%
where $\func{erf}\left( x\right) $ is the error function defined by, 
\begin{equation}
\func{erf}\left( x\right) \equiv \frac{2}{\sqrt{\pi }}\int_{0}^{x}\exp
\left( -t^{2}\right) dt.
\end{equation}%
Two immediate consequences or remarks follow. \emph{First,} defining a
temperature dependent length parameter $l_{T}$,%
\begin{equation}
l_{T}\equiv \sqrt{\frac{2}{\beta m\omega ^{2}}}\sim \sqrt{kT},
\end{equation}%
we find that $l_{T}\simeq r_{0}$ amounts to $T\approx T_{d}$, so a
reasonable interval of temperature is $T\in $ $(T_{l},T_{d})$ where $T_{l}$
is the liquefaction temperature which is determined by experiments or
theoretical calculations. In present study, we do not care much about the
precise value of this temperature $T_{l}$ which can be simply taken as zero
Kelvin without affecting our conclusion. 

\emph{Secondly,} the mean distance between two atoms are in fact of
temperature dependence via, 
\begin{equation}
\bar{r}=\frac{1}{Z_{1}^{vp}}\int_{0}^{\infty }r\exp (-\frac{1}{2}\beta
m\omega ^{2}\left( r-r_{0}\right) ^{2})dr=r_{0}+\frac{l_{T}\exp (-\left( 
\frac{r_{0}}{l_{T}}\right) ^{2})}{\sqrt{\pi }\left( \func{erf}\left( \frac{%
r_{0}}{l_{T}}\right) +1\right) }=r_{0}+r_{T},  \label{7}
\end{equation}%
and the mean of the square of the distance is, 
\begin{equation}
\overline{r^{2}}=\frac{1}{Z_{1}^{vp}}\int_{0}^{\infty r}r^{2}\exp (-\frac{1}{%
2}\beta m\omega ^{2}\left( r-r_{0}\right) ^{2})dr=r_{0}^{2}+r_{0}r_{T}+\frac{%
l_{T}^{2}}{2},
\end{equation}%
where, 
\begin{equation}
r_{T}\equiv \frac{l_{T}\exp (-\left( \frac{r_{0}}{l_{T}}\right) ^{2})}{\sqrt{%
\pi }\left( \func{erf}\left( \frac{r_{0}}{l_{T}}\right) +1\right) }.
\end{equation}%
The square root of the average value of standard deviation of the distance $%
r $, or the fluctuation of the distance, is defined by $\Delta r\equiv \sqrt{%
\overline{r^{2}}-\bar{r}^{2}}$ which is given by,\ 
\begin{equation}
\Delta r=\sqrt{\frac{l_{T}^{2}}{2}-r_{0}r_{T}-r_{T}^{2}}.
\end{equation}%
In whole temperature interval $T=\left( 0,T_{d}\right) $, the mean distance $%
\bar{r}$ remains almost unchanged as $r_{T}/r_{0}\simeq \left(
0,0.113\right) $, but\ it is not the case for the fluctuation of the
distance as $\Delta r/r_{0}\simeq \left( 0,0.612\right) $ from which we see
that as $T\rightarrow T_{d}$, the distance between two atoms fluctuates
dramatically. In other words, it is the fluctuation of the distance $\Delta
r $ rather than the elongated mean distance $\bar{r}$ cuts the bond as
temperature $T$ reaches $T_{d}$. Results of the fluctuation of the distance $%
\Delta r$ are plotted in FIG.1.

With the partition function $Z_{1}^{vp}$ (\ref{3}), the \emph{mean
vibrational potential energy} is, 
\begin{equation}
u^{vp}=-\frac{\partial }{\partial \beta }\ln Z_{1}^{vp}=\frac{1}{2}kT\left(
1-\frac{2x\exp (-x^{2})}{\sqrt{\pi }\left( 3-\func{erf}\left( x\right)
\right) }\right) ,
\end{equation}%
where $x\equiv r_{0}/l_{T}$, and $x\simeq 1$ when $T\simeq T_{d}$. When $%
T\ll T_{d}$, $x\gg 1$, the exponent function $\exp \left( -x^{2}\right) $
becomes much smaller, so the equipartition value is recovered,    
\begin{equation}
u^{vp}\simeq \frac{1}{2}kT.
\end{equation}%
The greatest deviation from the equipartition value $kT/2$ appears at $%
x\simeq 0.778$, at which $u^{vp}=0.394kT\prec 0.5kT$; and at $T\simeq T_{d}$%
, $u^{vp}=0.403kT\prec 0.5kT$. It is sufficiently to conclude that\ the
theorem of equipartition of energy is not applicable near the dissociation
temperature. Nevertheless, a more striking result is available from
examining the heat capacity $c^{vp}\equiv \partial u^{vp}/\partial T$, which
is simply, 
\begin{equation}
\frac{c^{vp}}{k}=\frac{1}{2}\left( 1-\frac{xe^{-x^{2}}\left( 2x^{2}+1\right) 
}{\sqrt{\pi }(1+\func{erf}(x))}-\frac{2e^{-2x^{2}}x^{2}}{\pi (1+\func{erf}%
(x))^{2}}\right) .  \label{last}
\end{equation}%
The standard result is $c^{vp}=0.5k$; and ours is smaller. The biggest
deviation occurs at $x\simeq 0.974\simeq 1$, i.e., $T\simeq T_{d}$, and we
have $c^{vp}=0.318k\prec 0.5k$. In other words, when the temperature is so
high as $T\simeq T_{d}$, i.e., $x\simeq 1$, which suffices to break the
bond, the theorem of equipartition of energy is seriously violated. The
result given by Eq. (\ref{last}) is plotted in FIG. 2. However, the theorem
of equipartition of energy is in fact valid in an ordinary condition which
is easily accessible in laboratories.

Take $O_{2}$ gas for instance. The basic data are listed in the following 
\cite{table}. The reduced mass of the oxygen molecule $m=1/2\ast 2.657\ast
10^{-26}$kg, and vibrational frequency $\omega =4.66\times 10^{13}$, and
bond length\ $r_{0}=148$ picometers. The dissociation temperature is $24548K$%
. At room temperature $T=300K$, $\beta =7.2\times 10^{22}J^{-1}$, we have $%
l_{T}=17.0$ picometers which lead to correction to the bond length by an
addition amount of small distance $r_{T}\simeq 5.82\ast 10^{-21}$
picometers, too bit to be meaningful. In opposite limit at the temperature $%
T=90.K$ at which the oxygen gas starts to liquefy \cite{table}, no deviation
from the theorem of equipartition of energy is observable as well.

The present study can be summarized in the following. A universal result is
identified for the vibrational heat capacity of the diatomic gas does not
obey the equipartition of energy at high temperature domain due to the
nonvanishing bond length. It is understandable in the following senses.
Because the theorem states that the mean value of every independent
quadratic term in the Hamiltonian is equal to $kT/2$, and only if all terms
in the Hamiltonian are quadratic then the mean energy is spread equally over
all degrees of freedom. This theorem says nothing about the possible
contributions from other non-quadratic terms in the Hamiltonian, and they
really have. One familiar example is the heat capacity resulted from
non-harmonic oscillations in crystal.

\begin{acknowledgments}
Q. H. is deeply grateful to Professor D. D. Yan at Beijing Normal University
and Professor L. Zhao at Nankai University for helpful comments. This work
is financially supported by National Natural Science Foundation of China
under Grant No. 11675051.
\end{acknowledgments}

\begin{figure}[ht]
\includegraphics[width=0.8\textwidth]{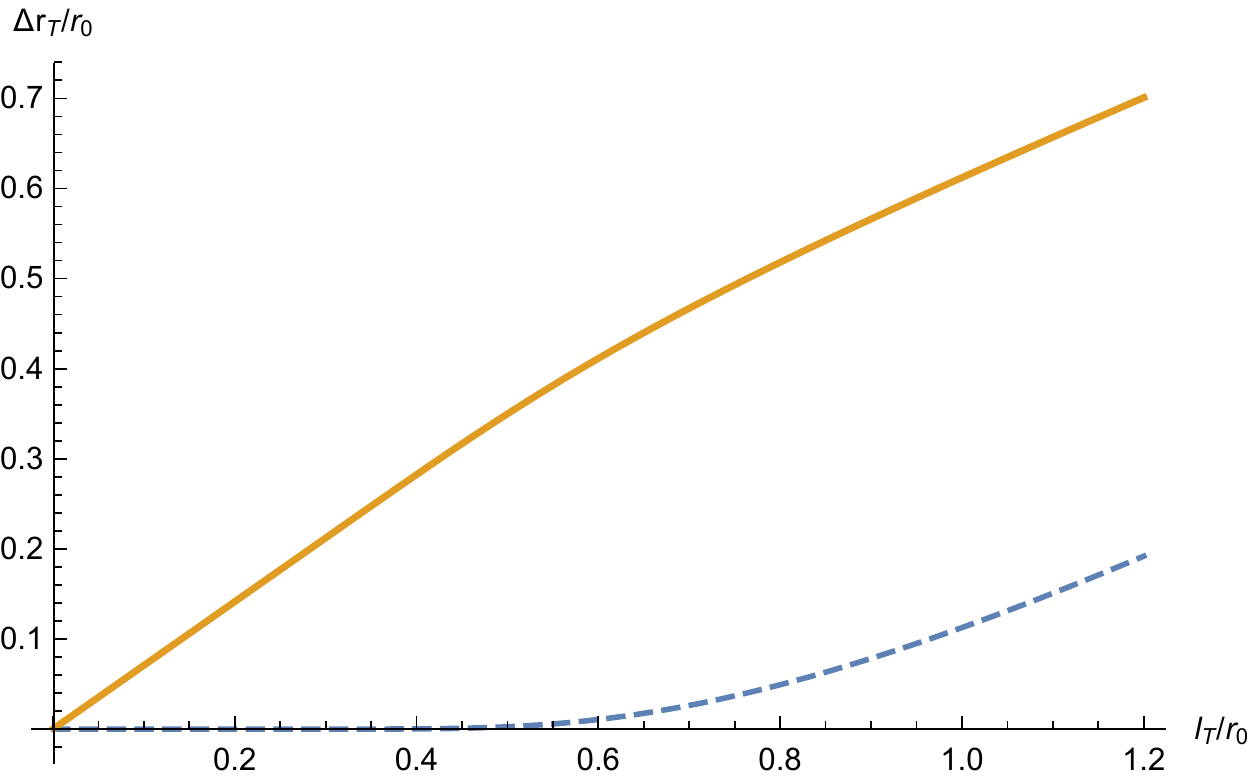}
\caption{The fluctuation of the distance between two atoms\ as $\Delta r\ $%
in unit bond length $r_{0}$ (solid line) and the mean distance $\bar{r}$ in
unit $r_{0}$ (dashed line, plotted as contrast) \textit{versus }the ratio%
\textit{\ }$l_{T}/r_{0}$ $\sim \protect\sqrt{kT}$. Results show that the
value of $\bar{r}$ changes a little, but that of $\Delta r$ changes
dramatically. Though these curves are meaningful only within temperature
interval $(T_{l},T_{d})$ in which $T_{l}$ is finite, we can also simply
assume $T_{l}\simeq 0$ at which the theorem of equipartition of energy
starts to apply.}
\label{Fig.1.}
\end{figure}

\begin{figure}[ht]
\includegraphics[width=0.8\textwidth]{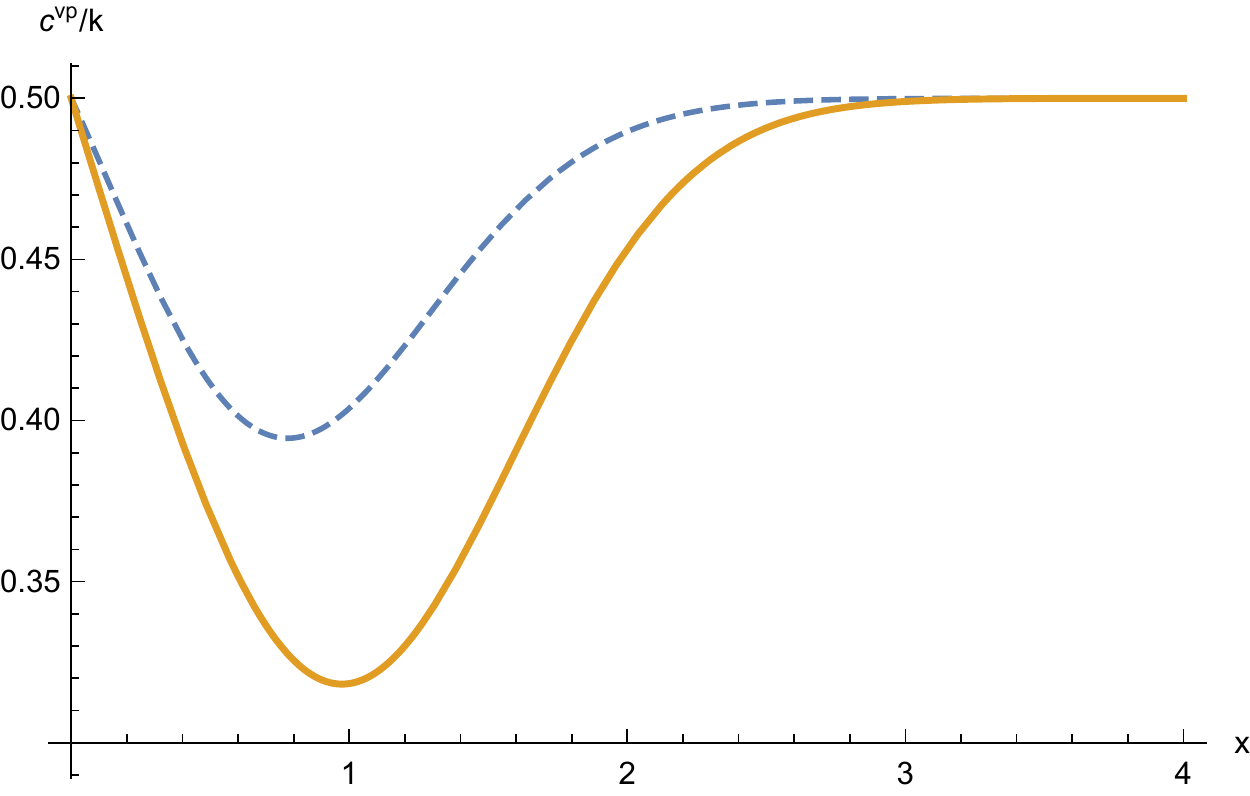}
\caption{The heat capacity $c^{vp}$ in unit Boltzmann constant $k$ from the
vibrational potential energy (solid line) and the \emph{mean vibrational
potential energy} $u^{vp}$ in unit $kT$ (dashed line, plotted as contrast) 
\textit{versus }the ratio\textit{\ }$x\equiv r_{0}/l_{T}\sim \protect\sqrt{%
\protect\beta }$. Results show that at relatively low temperature, the
theorem of equipartition of energy holds true, but becomes worse as $%
T\rightarrow T_{d}$. The same caution of the applicable temperature interval 
$(T_{l},T_{d})$ as that specified in FIG. 1 applies.}
\label{Fig.2.}
\end{figure}


\begin{thebibliography}{9}
\bibitem{1} R. K. Pathria, P. D. Beale, Statistical Mechanics, 3rd ed.,
(Oxford: Butterworth-Heinemann, 2011).

\bibitem{2} K. Huang, Statistical Mechanics, 2nd ed., (New York: Wiley,
1986).

\bibitem{3} M. Toda, R. Kubo, N. Saito, Statistical Physics I: Equilibrium
Statistical Mechanics, 2nd Ed., (Berlin: Sringer-Verlag, 2012).

\bibitem{4} F Reif. Fundamentals of Statistical and Thermal Physics, (New
York: McGraw-Hill, 1965).

\bibitem{5} L. D. Landau and E. M. Lifshitz, Statistical Physics (I), 3rd
ed., (Oxford: Pergamon, 1980).

\bibitem{table} J. G. Speight, Lange's Handbook of Chemistry, 16th Edition,
(New York: Mcgraw-Hill, 2005).
\end{thebibliography}
\end{document}